\documentstyle[preprint,aps]{revtex}
\newcommand{\be}{\begin{equation}}
\newcommand{\ee}{\end{equation}}
\newcommand{\eq}{\begin{equation}}
\newcommand{\en}{\end{equation}}
\newcommand{\bc}{\begin{center}}
\newcommand{\ec}{\end{center}}

\newcommand{\lsim}{\raisebox{0.3mm}{\em $\, <$} 
\hspace{-3.3mm} \raisebox{-1.8mm}{\em $\sim \,$}}
\newcommand{\gsim}{\raisebox{0.3mm}{\em $\, >$} 
\hspace{-3.3mm} \raisebox{-1.8mm}{\em $\sim \,$}}
\begin{document} 
\draft
\preprint{}
\title{
High Energy Cosmic Neutrinos and the Equivalence Principle 
}

\author{Hisakazu Minakata}
\address{Department of Physics, Tokyo Metropolitan University \\
Minami-Osawa, Hachioji, Tokyo 192-03, Japan}

\author{Alexei Yu. Smirnov}
\address{International Center for Theoretical Physics \\
P. O. Box 586, 34100 Trieste, Italy\\
and Institute for Nuclear Research, Russian Academy of Science \\ 
117312 Moscow, Russia}

\preprint{
\parbox{4cm}{
IC/96/9\\
TMUP-HEL-9601\\
hep-ph/9601311
}}
\maketitle
\begin{abstract}
Observation of the ultra-high energy neutrinos, in particular, 
detection of $\nu_{\tau}$ from cosmologically distant sources 
like active galactic nuclei (AGN) opens new possibilities to 
search for neutrino flavor conversion. We consider effects of 
violation of the equivalence principle (VEP) on propagation of 
these cosmic neutrinos. Two effects are studied:  
(1) the oscillations of neutrinos due to the VEP in the 
gravitational field of our Galaxy and in the intergalactic space, 
(2) the resonance flavor conversion driven by the gravitational 
potential of the AGN. 
We show that the ultra-high energies of the neutrinos as well as  
cosmological distances to the AGN, or strong AGN gravitational 
potential will allow one to improve the accuracy of test of the 
equivalence principle by 21 orders of magnitude for massless 
or degenerate neutrinos ($\Delta f \sim 10^{-41}$) and at least 
by 12 orders of magnitude for massive neutrinos 
($\Delta f \sim 10^{-28} \times (\Delta m^2/1 {\rm eV}^2)$).   
Experimental signatures of the transitions induced by the VEP 
are discussed. 
\end{abstract}
\newpage

\section{Introduction}

Cosmologically distant objects  such as the active galactic nuclei 
(AGN) can be intense sources of high-energy neutrinos \cite{AGN}. 
The flux of these cosmic neutrinos is flavor non-symmetric. 
Models predict that the $\tau$ - neutrino flux is at least 3 orders 
of magnitude smaller than the fluxes of the electron and muon 
neutrinos. This opens a unique possibility of searching for effects 
of neutrino flavor transitions over intergalactic distances.  

It was suggested recently \cite{LP} that large deep underwater and 
ice neutrino detectors will be able to identify the events induced by 
$\tau$ neutrinos with energies above 1 PeV. The $\tau$-neutrinos 
produce the characteristic double-bang events. The first bang comes 
from the charged current interaction of the $\tau$ neutrino and the 
second one originates from the hadronic decay of the $\tau$-lepton. 
At PeV energies the $\tau$-lepton tracks have typical lengths of the 
order of 100 m and the two bangs are clearly separable. Moreover, 
a selection criterion of greater energy of the second bang in 
comparison with the first one makes the events to be essentially 
background-free \cite{LP}, thereby guaranteeing the unambiguous 
detection of the $\tau$-neutrinos. 

It was estimated that the experiment is sensitive to the transition 
probabilities greater than 
\be 
\label{sensitivity}
P \geq (3 - 5) \cdot 10^{-3}. 
\ee 
The transition can be induced by the vacuum oscillations \cite{LP}. 
Apart from the appearance of the relatively large $\nu_{\tau}$-flux 
one expects also a modification of number of the 
$\nu_e$ and $\nu_{\mu}$ events. Cosmological distances to the AGN 
will allow one to probe the range of oscillation parameters: 
$\Delta m^2 >  10^{-15} {\rm eV^2}$, 
$\sin^2 2\theta > 2P \approx 10^{-2}$. 
In particular, strong effect is expected for the values of parameters 
needed to solve the atmospheric neutrino problem \cite{ATM}. 

In this paper we will consider the flavor transitions of the 
high-energy cosmic neutrinos induced by possible non-universality 
in the gravitational couplings of neutrinos. We show that the 
ultra-high neutrino energies and cosmological distances, or 
strong AGN gravitational potential, conspire to lead to 
great sensitivity in exploring the tiny non-universality.

The non-universality of the gravitational couplings of neutrinos 
can lead to the neutrino flavor oscillations \cite{Gasp}. 
Search for the neutrino oscillations in accelerator neutrino 
experiments and solar neutrino observation turns out to be a 
powerful tool of exploring possible violation of Einstein's 
equivalence principle \cite{HL,IMY,PHL,BNMB,MN,BKL,mina,panta}. 
Moreover, the oscillations and neutrino conversion
induced by violation of the equivalence principle supply viable
mechanism which could solve the solar and the atmospheric
neutrino problems simultaneously \cite{PHL,panta}. 

The sensitivities to non-universality achievable in these methods 
vary depending upon the experimental means, on the presence or 
absence of matter effect, and on values of  neutrino masses. 
Existing neutrino accelerator experiments restrict the 
non-universality parameter $\Delta f$ up to the level of 
$10^{-15}$ \cite{PHL,mina,panta}, assuming that the supercluster's 
gravitational potential is of order $10^{-5}$ (see below.) 
It will be possible to improve it in the planned long baseline 
accelerator experiments which can reach the sensitivity $10^{-18}$ 
\cite{IMY,PHL}.

In the case of massless (or degenerate) neutrinos studies of the 
solar neutrinos can be sensitive to $10^{-20}$ \cite{panta}. 
This limit corresponds to the vacuum long length oscillations. 
In the case of the resonant flavor conversion of massive neutrinos 
due to the matter effect in the solar interior the sensitivities 
are at most $10^{-15} - 10^{-16}$ \cite{MN,BKL,panta}.

It has also been discussed that for $\Delta f > 4 \cdot 10^{-14}$ 
the conversion induced by VEP leads  to substantial effect on 
supernova dynamics \cite{panta}. 
The arrival time difference between neutrinos and photons from 
SN1987a gives a modest bound of the order of $10^{-3}$ \cite{LKT}.  

We may conclude that by using any neutrino sources and any 
experimental devices so far considered it appears difficult to go 
far beyond the sensitivity of $\sim 10^{-20}$.

In this paper we will show that observation of the ultra-high energy 
neutrinos from cosmologically distant sources will allow one 
drastically improve the accuracy of testing the equivalence principle. 
The paper is organized as follows. In Sec. II we will consider 
the gravity-induced oscillations  of massless (or degenerate) 
neutrinos and estimate a sensitivity to  violation of the 
equivalence principle (VEP). In Sec. III the gravity effects in the 
presence of nonzero neutrino masses and vacuum mixing are discussed. 
In particular,  we describe the resonance flavor conversion driven by 
the gravitational potential. In Sec. IV we will apply the results to 
neutrinos from the  AGN. In Sec. V the experimental signatures of 
the VEP effects are considered. In Sec. VI we summarize the results.  
  

\section{Gravitationally induced oscillations 
of neutrinos from the AGN}

Let us restrict ourselves to the two-flavor case,  
$(\nu_{\mu}, \nu_{\tau})$,  for simplicity. 
According to the hypothesis of violation of the equivalence principle 
(VEP) \cite{Gasp} the flavor eigenstates 
$\nu_{\mu}$ and $\nu_{\tau}$ 
are the mixtures of the gravity eigenstates, 
$\nu_{2g}$, $\nu_{3g}$,    
whose gravitational couplings $f_{2}G$ and $f_{3}G$,  
where $G$ is the  Newton constant, 
are different  $f_2 \neq f_3$. Evidently,   
$f_i \neq 1$ at least for one neutrino.   
Introducing gravitational mixing angle,  
$\theta_g$, one can write 
\be
\nu_{\mu} = \cos \theta_g \nu_{2g} + \sin \theta_g \nu_{3g}, \ \    
\nu_{\tau} = - \sin \theta_g  \nu_{2g} + \cos \theta_g \nu_{3g} \ .  
\ee  
The non-universality of the  gravitational couplings 
can be parametrized as 
\begin{equation}
\Delta f = {f_3 - f_2   \over {1 \over 2}(f_3 + f_2)} \ .
\end{equation} 
The $\nu_{2g}$ and  $\nu_{3g}$     
 neutrinos feel gravitational fields with 
slightly different strengths. This leads to a 
difference in the energies 
of the eigenstates (i.e. to the level energy splitting):  
\begin{equation}
\label{poten}
V_g  \equiv {1 \over 2}\Delta f E \Phi(x),  
\end{equation}
where $E$ is the energy of neutrino,  and $\Phi(x) = MG/r$ is the 
gravitational potential at distance  $r$ from an object of mass $M$ 
in the Keplerian approximation. The energy level splitting induces 
a relative phase difference between 
wave functions of $\nu_{2g}$ and  $\nu_{3g}$     
which results in neutrino flavor oscillations  
in the same fashion as in the 
mass-induced case. 

Let us first suggest that neutrinos are massless or have equal masses.  
As we will show later the matter effect is negligibly small. 
In this case the propagation of neutrinos has a character of 
oscillations with the depth fixed by $\theta_g$ 
and with the length, $l_g$,  
determined by $V_g$:   
\begin{equation}
\label{length}
l_g = \frac{2\pi} {\Delta f E \Phi(x)}\ .   
\end{equation} 
The oscillation probability can be written as 
\begin{equation}
\label{probab}
P = \sin^2 2\theta_g \sin^2 \left(\Delta f E 
\int {\Phi(x)}dx \right)\ . 
\end{equation} 
The distinctive feature of (\ref{length}) is that the oscillation 
length is inversely proportional to the neutrino energy, 
in contrast with the vacuum oscillation length which is directly 
proportional to $E$. Therefore the  sensitivity to $\Delta f$ 
increases with the neutrino energy and with the path-length in 
the gravitational field. Both factors are present for neutrinos 
from the AGN. The AGN are believed to produce high energy  
neutrinos with spectrum extended to $E \sim 10$ PeV, 
\cite{AGN} and typical distances from our Galaxy to the AGN are 
$L_{AGN} \sim 100$ Mpc.\\ 

Let us estimate a sensitivity to $\Delta f$. For this purpose 
we should find the integral $I = \int {\Phi(x)}dx$ along the 
neutrino trajectory which appears in (\ref{probab}). 
The intergal has three contributions:
\be
I = I_{AGN} + I_{IG} + I_G, \ \ \ \ 
I_i \equiv \int {\Phi_i(x)}dx, \ \ (i = AGN,~ IG,~ G) \ ,  
\ee
where  
$\Phi(x)_{AGN}$,    
$\Phi(x)_{IG}$,  and 
$\Phi(x)_G$ 
are the potentials created by the AGN itself, by all bodies in   
the intergalactic space and by our Galaxy, 
respectively. 

We get $I_{AGN}(r) \approx -{1 \over 2} R_S \log {(r/R_e)}$ for a 
radial trajectory, where $R_S$ is the Schwarzschild radius of AGN:  
$R_S \simeq 3 \times 10^{11}(M_{AGN}/10^8 M_{\odot})$ m and 
$R_e \approx (10 - 10^{2}) R_s$ is the radius of the neutrino 
emission region. Using $M_{AGN} \sim 10^{8} M_{\odot}$ as a typical 
mass of the AGN,  we find 
 $R_S \sim 3 \times 10^{11}$ m and $R_e \sim 10^{13}$ m ,  
and consequently 
\be
I_{AGN} \sim 10^{-10} {\rm Mpc}.
\ee

In the intergalactic space the effect is dominated by the   
gravitational field of the so called Great Attractor \cite{LB}. 
This supercluster is located at the distance  (43.5 $\pm$ 3.5) 
$h_0^{-1}$ Mpc from the Earth, where $h_0$ is in the range 0.5-1.0. 
The  mass of the Great Attractor is about 
$M_{sc} \sim 3\times10^{16}h_0^{-1}M_{\odot}$, where 
$M_{\odot}$ is the solar mass. 
\cite{Kenyon}. In the Keplerian approximation the gravitational 
potential of this supercluster can be estimated as 
$\Phi(R) = 
- 5.2 \times 10^{-6}(R/100 Mpc)^{-1}(M_{sc}/10^{16} M_{\odot})$. 
(Note that the weak-field approximation still applies). 
Then the  integral $I_{IG}$ satisfies the inequality 
\be
|I_{IG}| \gsim 5.2 \times 10^{-4}
\left ( \frac{L}{100 {\rm Mpc}} \right)
\left( \frac{M_{sc}}{10^{16} M_{\odot}}\right)\  {\rm Mpc} 
\ee
for any trajectory of length $L$ within radius of 100 Mpc. 

The contribution of our Galaxy equals to 
$I_G = - GM_{G} \log {(L_{AGN}/r)}$ 
for radial trajectory. Using $M_G \sim 10^{11} M_{\odot}$   
and $r = 10$ kpc for the mass and the radius of the Galaxy, 
respectively, and $L_{AGN} = 100$ Mpc we get  
$I_G \simeq - 10^{-7}$ Mpc.  

The supercluster gravitational effect dominates: $I \approx I_{IG}$. 
The reason is that the path-length in the gravitational field 
tends to cancel the inverse distance dependence of the Keplerian 
potential. Therefore, the oscillation probability is essentially 
governed by the mass of source of the gravitational field and 
the mass relation  $M_{sc} \gg M_G \gg M_{AGN}$ 
leads to $I_{sc} \gg I_{G} \gg I_{AGN}$.   

It is straightforward to make an order-of-magnitude estimation of the 
sensitivity.  For  $E$= 1 PeV and $I_{IG}$ 
(at $\Phi = 10^{-5}$, and the distance $L = 100$ Mpc) one gets
\begin{equation}
|I_{IG} E| > |\Phi E L| =  1.5 \times 10^{41} 
\label{argsine}
\end{equation}
which means according to (\ref{probab}) that the phase of 
oscillation of order unity will be obtained for 
$\Delta f > 10^{-41}$. 

If the great attractor is a fake object, then the dominant effect 
would be due to the gravitational field of our Galaxy: 
$I \approx I_G$. In this case the expected sensitivity 
to $\Delta f$ is $\sim 10^{-37}$ which still implies an improvement 
by more than 17 orders of magnitude. This $\Delta f$ can be 
considered as the conservative estimation of the sensitivity.\\  

Let us show that in spite of cosmological distances 
the matter effect on the neutrino conversion 
can be neglected. Consider the $\nu_{e} - \nu_{\tau}$ 
system for which matter effect appears in the 
first order in the weak interactions. The cosmological baryon 
density estimated from the  nucleosynthesis is  
$\rho_B \sim 10^{-31}$ g/cm$^3$. This gives the width of matter  
in the intergalactic space: 
$d_{IG} \equiv \rho_B \cdot L_{AGN} \sim 3 \cdot 10^{-5}$ g/cm$^{2}$.  
According to  spheroid-dark corona models \cite{Schmidt}  
the matter density of our Galaxy is 
$\rho \sim (1-10) \times 10^{-25}$ g/cm$^3$. 
This leads to the width 
$d_{G}  \sim 3 \cdot (1-10) \times 10^{-3}$ g/cm$^{2}$. 
Finally, the width of matter crossed by neutrinos 
in the AGN is estimated as  
$d_{AGN}  \sim  (10^{-2} - 10^{-1})$ g/cm$^{2}$ \cite{Xray}. 
Thus total width,  
$d_{total} \approx d_{AGN}  \sim  (10^{-2} - 10^{-1})$ g/cm$^{2}$,  
is much smaller than the effective width  
$d_0 \equiv  \sqrt {2}\pi m_N /G_F 
\approx 2 \cdot 10^9$  g/cm$^{2}$ needed for appreciable matter 
effect. For $\nu_{\mu} - \nu_{\tau}$ channel the matter effect 
appears in high order of perturbation theory and the required 
effective width is even larger. \\ 


\section{Gravitationally induced transitions in the presence of 
neutrino masses} 

Most probably neutrinos are massive and mixed. The gravitational 
effects themselves can generate via the nonrenormalizable 
interactions the neutrino masses of the order 
$v^2/ M_P \sim 10^{-5}$ eV \cite{ABS}, 
where $v$ is the electroweak scale and $M_P$ is the Planck mass. 
Moreover, there are some hints from solar, atmospheric, cosmological 
as well as accelerator data that neutrino masses are even larger than 
that value. Forthcoming experiments will be able to check the hints.  
In this connection we will consider the gravitational effects in the 
presence of neutrino masses and mixing, assuming that the 
latter will be determined from the forthcoming experiments.

In the presence of the vacuum and gravity mixing the effective 
Hamiltonian of neutrino system in the flavor basis 
$\nu = (\nu_{\mu}, \nu_{\tau})$ 
reads \cite{Gasp} (for notation we follow \cite{MN}): 
\be 
\label{hamilt}
H(x) = \delta \left(
\matrix{
-c & s \cr
 s & c \cr}
\right)
+ V_g \left(
\matrix{
-c_g & s_g \cr
 s_g & c_g \cr
}
\right) , 
\ee
where $\delta \equiv \Delta m^2 /4E$ with 
$\Delta m^2 \equiv {m_3}^2 - {m_2}^2$, $c \equiv \cos 2 \theta$, 
and $c_g \equiv \cos 2 \theta_g$, {\it etc}.. As we have shown 
in Sec. II the matter effects can be neglected. For antineutrinos 
the VEP may differ from that for neutrinos: 
$\Delta \bar{f} \neq \Delta f$, so that one may expect a 
difference of the effects in neutrino and antineutrino channels. 
We assume at the moment that all the parameters of the Hamiltonian 
(\ref{hamilt}) are real. In general, the complex phase can be 
introduced in (\ref{hamilt}) which has some physical consequences 
\cite{panta}. We will discuss effects of the phase at the end of 
this section.\\

According to  (\ref{hamilt})  
the mixing angle in medium, $\theta_m$, is fixed by   
\be
\label{mixing}
\tan 2 \theta_m = \frac
{s \delta + V_g s_g}
{c \delta + V_g c_g}  \ . 
\ee
Evidently, for $V_g \gg \delta$ the mixing is determined by gravity: 
$\theta_m \approx  \theta_g$. And for $V_g \ll \delta$ 
the mixing angle equals the vacuum angle: 
$\theta_m \approx  \theta$. 

As follows from  (\ref{mixing}) the mixing angle is zero at 
\be
\label{zero}
V_g^0 = - \delta \frac{s}{s_g}  \ . 
\ee

If $\theta_g =  \theta$ 
one has from (\ref{mixing}) $\tan 2\theta_m =  \tan 2\theta$. 
In this case the evolution of neutrino state is reduced to  
oscillations with the constant depth, $\sin^2 2\theta$,  
and the oscillation length 
\be
\l_{\nu} = \frac{2 \pi}{\delta + V_g} \ \ .
\ee 
The phases due to mass difference and the gravitational 
effect add up. The gravitational phase dominates if $V_g > \delta$, 
or explicitly, if $\Delta f > \Delta m^2 /(2 E^2 \Phi)$. 
For supercluster potential, $\Phi_{sc} \sim 10^{-5}$,  
and $E = 1$ PeV we get from this inequality  
$\Delta f > 5 \times 10^{-36}$,  $5 \times 10^{-28}$, 
$5 \times 10^{-25}$ for 
$\Delta m^2 = 10^{-10}, 10^{-2}, 10$ eV$^2$, 
respectively. However, in all these cases the oscillations 
are averaged and even if $\Delta m^2$ will be known it is  
impossible to identify the gravity effects.\\ 

If $\theta \neq  \theta_g$,  the Hamiltonian (\ref{hamilt}) can 
lead to the resonance flavor conversion due to  change of $V_g$ 
with distance, i.e. to the gravity version of the 
Mikheyev-Smirnov-Wolfenstein (MSW) mechanism \cite{MSW}.   
Note that apart from a brief remark in \cite{MN} the consideration 
in the existing literature is restricted to the level crossing 
driven by matter density change with distance. Here we describe 
the level crossing driven by gravitational potential change 
rather than matter density change.

As follows from (\ref{hamilt}) the resonance condition 
(condition for maximal mixing) is 
\be 
V_g = - \frac{c \delta}{c_g} \ , 
\label{resonance}
\ee 
and the resonance value of the potential  equals  
\be
\label{res}
\Phi_R = 
- \frac{\Delta m^2}{2 E^2 \Delta f}
\frac{\cos 2 \theta}{\cos 2 \theta_g} \ . 
\ee 
Correspondingly, $V_g$  in the resonance 
is $V_g^R = \Delta f E \Phi_R/2$. 

The adiabaticity condition in the resonance reads \cite{MN,BKL}
\be
\label{adiab}
\left| \frac {(\delta \sin2\theta + V_g \sin2\theta_g)^2 } 
{\frac{d}{dr}V_g \cos2\theta_g} \right|_{resonance} \gg 1.
\ee
This condition simplifies under the Keplerian approximation. 
Substituting in (\ref{adiab}) $d V_g/dr = - V_g/r$ and 
using the resonance condition (\ref{resonance}) we can 
rewrite (\ref{adiab}) as 
\be
\label{adiabaticity} 
(\tan2\theta - \tan2\theta_g)^2 \cos2\theta_g 
\gg \frac{1}{r_R \delta} \ , 
\ee
where $r_R$ is the radius at which the resonance condition is 
fulfilled: $r_R = r_R(\Delta f)$. 

The conditions (\ref{resonance}, \ref{adiabaticity}) 
determine the sensitivity regions of parameters 
$\Delta f$, $\theta_g$. Indeed, for fixed $\Delta m^2$ and 
$c \approx c_g \approx 1$ the minimal and maximal values of the 
potential $\Phi$ determine via the resonance condition the range of 
$\Delta f$ for which the resonance neutrino conversion may take place.
Maximal value of $r_R$ at which the resonance condition is fulfilled 
gives the lower bound on mixing angles through the adiabaticity
condition (\ref{adiabaticity}). If the adiabaticity condition is 
satisfied with vacuum mixing angle alone, i.e., 
$\sin^2 2\theta \approx \tan^2 2\theta \gg (r_R \delta)^{-1}$,  
then $\theta_g$ can be arbitrarily small. In this case the role of 
the gravitational effect is just to split levels. On the contrary, 
for $\theta = 0$ one has the lower bound on the gravitational mixing
\be
\label{adiabangle}
\sin^2 2\theta_g \gg \frac{1}{r_R \delta}\ .
\ee

For fixed mixing angles the adiabaticity condition can be rewritten 
as the lower bound on $\Delta f$. Indeed, substituting $\delta$ in 
the resonance condition (\ref{resonance}) into (\ref{adiabaticity}) 
we find 
\be
\label{adi} 
\Delta f \gg \frac{2 \times 10^{-33}}
{(\tan2\theta - \tan2\theta_g)^2 \cos2\theta_g}
\left(\frac{E}{1 {\rm PeV}} \right)^{-1}
\left(\frac{M_{AGN}}{10^8 M_{\odot}} \right)^{-1} \ .
\ee
\\

If the adiabaticity condition is fulfilled, then the transition 
probability is determined by the initial and final values of 
mixing angle:   
\be
\label{transition1}
P_a = \frac{1}{2}\left(1 - \cos 2 \theta_{mi} \cos 2 
\theta_{mf} \right)\ .   
\ee
In this connection let us consider a dependence of mixing angle 
$\theta_m$ on the potential $\Phi$ or level splitting 
$V_g = \Delta f \Phi E/2$. The angle $\theta_m$ as the function 
of $\Phi$ crucially depends on the sign of $\Delta m^2 \Delta f$, 
and on whether $\theta_g > \theta$ or $\theta_g <  \theta$. 
We focus first on the resonant channel, $\Delta m^2 \Delta f > 0$.   

\noindent
1). $\theta_g < \theta$. In this case one has $|V_g^0| > |V_g^R|$,  
i.e. the value of potential which corresponds to 
zero mixing (\ref{zero}) is bigger than the resonance value.    
For the initial splitting  $|V_g^i| \gg |V_g^R|$, the gravity mixing 
dominates and $\theta_m \approx \theta_g + \pi/2$. 
With diminishing $|V_g|$, the angle $\theta_m$ decreases 
and mixing becomes zero ($\theta_m =  \pi/2$)  
at $V_g = V_g^0$; then $\theta_m$ crosses the resonance value, 
$\theta_m = \pi/4$,  
and for $|V_g| \ll |\delta|$  approaches  $\theta_m = \theta$.   

\noindent
2). $\theta_g > \theta$.  Now $|V_g^0| <  |V_g^R|$. At 
$|V_g^i| \gg |V_g^R|$ one has  $\theta_m \approx \theta_g - \pi/2$.  
With diminishing $|V_g|$ the angle $\theta_m$ increases and crosses 
resonance value 
$\theta_m = - \pi/4$. At $V_g = V_g^0$ the angle $\theta_m$ 
vanishes so that $\sin^2 2\theta_m = 0$, and then 
$\theta_m$ approaches the vacuum value $\theta$.  

In the nonresonant channel, movement of the angle $\theta_m$ 
is simpler. Under the same variation of $V_g$ as above it 
starts from $\theta_m = \theta_g$ and ends up with $\theta$ 
without crossing the points of 
zero mixing and resonance irrespective of the relative 
magnitudes of $\theta_g$ and $\theta$.\\

Suppose that the initial potential (the potential at the 
production point) is much larger than the one at resonance, 
$\Phi_i \gg \Phi_R$, and the final potential is much smaller 
than the value at resonance, $\Phi_f \ll \Phi_R$. In this case  
the transition probability in the adiabatic approximation 
(\ref{transition1}) becomes: 
\be
\label{transition}
P_a = \frac{1}{2}\left(1 \pm \cos 2 \theta_g \cos 2 
\theta \right) ,  
\ee
where the plus sign is for the resonance channels 
($\Delta m^2 \Delta f > 0$), and the minus sign is for the 
non-resonant channels ($\Delta m^2 \Delta f < 0$). 

Let us mark one interesting feature related to zero mixing at 
$V_g = V_g^0$ (\ref{zero}). If the initial (final) potential is 
such that 
$V_g = V_g^0$  for $\theta_g < \theta$ ($\theta_g >  \theta$),  
then the transition probability in the resonant channel reduces to 
$P = \cos^2\theta$ ($P = \cos^2\theta_g$) as in the case of 
flavor conversion in matter.\\

Let us discuss finally effects of possible complex phases  
in the Hamiltonian  (\ref{hamilt}). After redefinition of the 
neutrino wave function only one phase survive which can be put in 
the gravitational term of the Hamiltonian (\ref{hamilt}) \cite{panta}: 
\be 
\label{hamiltg}
V_g \left(
\matrix{
-c_g & s_g e^{-i2\alpha} \cr
 s_g e^{i2\alpha} & c_g \cr
}
\right) \ . 
\ee
The angle $\alpha$ is the relative phase of the vacuum and gravitational 
contributions to the mixing (nondiagonal elements), and it is present 
when both of these contributions exist. 
Let us consider physical effects of the phase.

The Hamiltonian (\ref{hamilt}) can be rewritten as 
\be 
\label{hamiltA}
H(x) = 
\left(
\matrix{
- c \delta -  V_g c_g  & \rho_{\alpha} e^{-i2\psi} \cr
\rho_{\alpha} e^{i2\psi} & c \delta + V_g c_g  
                \cr}
\right) \ ,
\ee
where 
\be 
\label{rho}
\rho_{\alpha} = \left|\delta s + V_g s_g e^{-i2\alpha}\right|
= V_g s_g \sqrt{\xi^2 + 2 \xi \cos 2\alpha +1} \ , 
\ee
with $\xi \equiv {\delta s}/ {V_g s_g}$,  
and 
\be
\label{tan}
\tan 2\psi = \frac{V_g s_g \sin 2\alpha}
{\delta s + V_g s_g \cos 2 \alpha} \ .
\ee
Additional redefinition of the fields 
\be
\nu_{\mu}' = e^{i\psi} \nu_{\mu} \ , \ \ 
\nu_{\tau}' = e^{-i\psi} \nu_{\tau} , \ \ 
\ee
leads to the evolution equation for 
$(\nu_{\mu}',   \nu_{\tau}')$ with the Hamiltonian 
\be 
\label{hamilt..}
H(x) = 
\left(
\matrix{
- c \delta - V_g c_g - \dot{\psi}      & \rho_{\alpha} \cr
\rho_{\alpha}    & c \delta + V_g c_g + \dot{\psi}  \cr}
\right) \  ,
\ee
where $\dot{\psi} \equiv d\psi/dr$.\\

From this we get the following consequences.  

\noindent
1). There is no complex phases in the Hamiltonian (\ref{hamilt..}).  
Thus, the presence of the phase $\alpha$ does not lead to the 
$CP-$ or $T-$ violating effects in the two generation case, 
as it was expected from the beginning.   

\noindent
2). The resonance condition is modified: 
\be
c\delta + V_g c_g + \dot{\psi} = 0 \ . 
\ee
The shift of the resonance, $\dot{\psi}$,  
is absent in the case of constant gravitational potential for which 
$\dot{\psi} = 0$. 
If the adiabaticity condition is fulfilled in the absence of 
$\alpha$ then the shift of the resonance position is negligibly 
small. Indeed, for a maximal phase $\sin^2 2\alpha = 1$, 
we find from (\ref{tan}) that 
$ 
\dot{\psi} < \frac{\dot{V_g}}{4V_g}
$
and then after suitable modification including the effect of phase,  
the adiabaticity condition leads to  
$ 
\dot{\psi} \ll V_g s_g
$.
The latter condition means that the shift of the resonance is much 
smaller than the width of the resonance. 

\noindent
3). Nonzero phase $\alpha$ modifies the dependence of the mixing 
parameter $\sin^2 2\theta_m$ on $V_g$. The mixing is proportional 
to the nondiagonal element of the Hamiltonian (\ref{hamilt..}) 
$\rho_{\alpha}$. For $\xi \gg 1$ or $\xi \ll 1$
it follows from (\ref{rho}) that 
$\rho_{\alpha} \approx \rho_0$, where $\rho_0$ corresponds to the 
case of zero $\alpha$. That is, 
the effect of $\alpha$ is small when the gravity or vacuum effect 
dominate. 

The strongest influence of $\alpha$ is for $\xi = -1$ which 
corresponds to  the point of zero mixing:  $V_g = V_0$.   
In fact, the phase remove zero mixing at $V_0$.  

\noindent
4). If $\xi \cos 2\theta > 0$, then $\rho_{\alpha} < \rho_0$; 
the phase leads to decrease of mixing and to narrower resonance peak. 
For $\xi \cos 2\theta < 0$ we get $\rho_{\alpha} > \rho_0$, 
i.e., the phase enhances mixing.  

\noindent
5). The presence of phase affects  also  the adiabaticity 
condition. However, it can be shown that if the adiabaticity is 
fulfilled at $\alpha = 0$, this influence is weak apart from 
some exceptional values of the phase. 

Thus we see that in many physically interesting situation the 
role of the complex phase is quite small even if $\alpha \sim O(1)$, 
and in what follows we will discuss the case $\alpha = 0$.


\section{Flavor transition of the AGN neutrinos}

Let us consider the flavor transitions of neutrinos from the AGN 
using the results of Sec. III. Following the scenarios described 
in \cite{AGN}, we assume that neutrinos are produced within the 
region located at the distance $R_e = (10-100) R_S$ from the center 
of AGN, where $R_S$ is the Schwarzschild radius:  
$R_S \simeq 3 \times 10^{11}(M_{AGN}/10^8 M_{\odot})$ m.   
For radii larger than the neutrino production point   
we may use the Keplerian approximation for the potential of the AGN.    
The total potential probed by neutrinos on the way to the Earth is 
\be
\label{AGNpotential}
\Phi(r) \approx \Phi_{AGN}(r) + \Phi_{IG} = 
\Phi_{AGN}^0 
\left( \frac{R_e}{r} \right)  + \Phi_{IG} \ .
\ee
Here, 
\be
\Phi_{AGN}^0  \simeq 
- 5 \times 10^{-3} \left( \frac{M_{AGN}}{10^8 M_{\odot}} \right) 
\ee
is the AGN potential at the neutrino production point and, for 
simplicity, we take the potential in the intergalactic space to 
be constant: 
$\Phi_{IG} = 10^{-5}$.  
Therefore, at the neutrino production point the potential 
$\Phi_{AGN}^0$ dominates over the supercluster and the galactic 
potentials.  For AGN located at 100 Mpc from us it is about 3 and 
7 orders of magnitude larger than the potentials of the great 
attractor and the Milky Way Galaxy, respectively. In what follows 
we neglect the potential of our Galaxy. \\ 

For fixed $\Delta f$ the dependence of the transition 
probability on the neutrino energy (or $E^2/\Delta m^2$) is 
the following. For small energies the mass-induced vacuum 
oscillation effect dominates and 
\be
\label{small}
P \approx \frac{1}{2} \sin ^2 2\theta \ \ \ \ \   {\rm for}  \ \ 
\frac{E^2}{\Delta m^2} < \frac{1}{2 \Phi_{AGN}^0 \Delta f} \ . 
\ee
For larger energies the transition probability equals 
$P_a$ in (\ref{transition}), if the adiabaticity condition 
is fulfilled: 
\be
\label{conversion}
P \approx P_a \ \  \ \  {\rm  for} \ \ 
\frac{E^2}{\Delta m^2} >
\frac{1}{2 \Phi_{AGN}^0 \Delta f} \ .
\ee
Moreover, in the region where AGN potential dominates 
over intergalactic potential the resonance conversion takes place. 
This corresponds to  
\be
\label{resconversion}
\frac{1}{2 \Phi_{AGN}^0 \Delta f} 
 < \frac{E^2}{\Delta m^2} < 
\frac{1}{2 \Phi_{IG} \Delta f} \ . 
\ee
For these energies the transition probability is larger than 1/2, 
and can be close to 1. The weaker the potential of the supercluster 
the wider the resonance-effective region of parameters. Moreover, 
if the $\Phi_{IG} < 10^{-7}$ the resonance conversion may take place 
in the gravitational field of our Galaxy too. 

For higher energies the mass splitting can be neglected and the 
dominant effect is the one due to gravitationally induced 
oscillations:  
\be
\label{large}
P \approx \frac{1}{2} \sin ^2 2\theta_g \ \ {\rm  for}  \ \  
\frac{E^2}{\Delta m^2} \gg  
\frac{1}{2 \Phi_{IG} \Delta f} \ . 
\ee
\\
Once the adiabaticity condition is satisfied in the AGN gravitational 
field the transition probability (\ref{transition}) applies also 
to the nonresonant channels.

The resonance flavor conversion is effective thanks 
to the fact that the mass of the AGN is concentrated 
mainly in a small region of space ($\sim 10^{-5}$ kpc) of the 
central black hole. Since our Galaxy is more massive than the 
typical AGN one might expect that neutrinos converted at the 
AGN could be reconverted by our Galaxy's gravitational field. 
It does not occur for $\Phi_{IG} \sim 10^{-5}$: the  
gravitational potential of the Galaxy is at most 
$\Phi_G \sim 10^{-7}$ because its mass is distributed over 
the $\sim 10$ kpc region.   

Let us find the sensitivity to $\Delta f$ assuming that future 
underwater/ice experiments will be sensitive to the flavor 
transition with probability as small as in (\ref{sensitivity}) 
\cite{LP}. We can distinguish three ranges of $\Delta f$ 
corresponding to three energy regions defined in 
(\ref{small}), (\ref{resconversion}) and (\ref{large}).\\ 
\noindent
{\it Region I}:
\be
\label{vacuum}
\Delta f < \Delta f_{AGN} \ , 
\ee
where 
\be
\label{fAGN}
\Delta f_{AGN} = 10^{-30} \times 
\left(\frac{\Delta m^2}{10^{-2}{\rm eV}^2} \right)
\left( \frac{\bar{E}}{1 \mbox{PeV}} \right)^{-2}
\left( \frac{R_e}{100 R_S} \right)
\left( \frac{M_{AGN}}{10^8 M_{\odot}} \right)^{-1}.
\ee
Here $\bar{E}$ is the average energy of the detected neutrinos. 
The gravity does not play any role and the effect of mass-induced 
vacuum oscillation dominates.

\noindent
{\it Region II}: 
\be
\label{fresonance}
\Delta f_{AGN} < \Delta f < \Delta f_{IG} \ , 
\ee
where 
\be
\label{fIG}
\Delta f_{IG} = 5 \cdot 10^{-28} \times 
\left( \frac{\Delta m^2}{10^{-2}eV^2} \right)
\left( \frac{\bar{E}}{1 \mbox{PeV}} \right)^{-2}
\left( \frac{10^{-5}}{\Phi_{IG}} \right) \ .
\ee
In this region neutrinos undergo resonance conversion driven 
by the  potential of the AGN, and the transition probability can be 
close to 1,  if both $\theta$ and $\theta_g$ are small. 

Suppose that the factor 
$(\tan 2\theta - \tan 2 \theta_g)^2 \cos 2 \theta_g$ 
in the adiabaticity condition (\ref{adi}) is of order unity.
Then the adiabaticity condition is satisfied for 
$\Delta f \gg 2 \times 10^{-33}$. 
Comparing this inequality with (\ref{fIG}) we find that for 
$\Delta m^2 \gsim 10^{-4}$ eV$^2$ the adiabaticity holds in 
whole region where the resonance condition is met. 
If $10^{-7}$ eV$^2$ $\lsim \Delta m^2 \lsim 10^{-4}$ eV$^2$ 
the adiabaticity region partially overlaps with the resonance region.
If $\Delta m^2 \lsim 10^{-7}$ eV$^2$ there is no overlapping between 
the two regions and the resonance conversion is not efficient. 

\noindent
{\it Region III}: 
\be
\label{gravdom}
\Delta f \gg \Delta f_{IG} \ . 
\ee
Here gravitational effect dominates and neutrinos  
oscillate due to the VEP. The probability 
converges to $P = \frac{1}{2} \sin^2 2\theta_g$.


\section{Observational signature of gravity effects}

The underwater/ice detectors will be able to measure the ratio 
of number of events induced by 
$\nu_{\tau}$ and $\nu_{\mu}$, $\nu_\tau/\nu_\mu$,  
as well as the ratio of $\nu_{e}$ and $\nu_{\mu}$ events,  
$\nu_e/\nu_\mu$.

If neutrinos are massless or degenerate, 
the  excess  
of the $\tau$ neutrino signal in underwater installations 
will testify  
for the neutrino flavor transition induced by the VEP. 
As we discussed in sect. 2 the appropriate range of parameters 
is  $\Delta f > 10^{-41}$ and $\sin^2 2\theta_g \gsim 0.01$. 

How can one prove this? 

Evidently, the anisotropy of $\nu_\tau/\nu_\mu$ and $\nu_e/\nu_\mu$   
ratios correlated to the position of the supercluster and/or the 
center of our Galaxy could be a signature of the gravitational 
origin of the excess. Let us compare the gravitational effect in 
the direction towards and away from the supercluster. Consider 
the AGN's at the distance $R$ from the center of supercluster in 
the cone directed from the Earth toward and outward the supercluster. 
Neutrinos from such AGN will acquire, on the way to the Earth, 
the gravitational phases 
\be
\phi_g^{toward} = \Delta \phi + \phi\ , \ \  
\phi_g^{away}  = \phi \ , 
\ee
where 
\be
\Delta \phi = 
\Delta f E M_{sc} G \left[2 \ln\frac{d}{r_0} +1\right]\ , \ \  
\phi = \Delta f E M_{sc} G \left[\ln \frac{R}{d}\right]  \ .
\label{toward}
\ee
Here $d$ is the distance from the Earth to the supercluster and 
$r_0$ is the size of the supercluster. The difference of phases is 
described by $\Delta \phi$. The effect is determined by integration 
over $R$. (If we neglect the expansion of the Universe the flux of 
neutrinos from different $R$ will be the same in the case of 
uniform distribution of the AGN). The anisotropy will be observable 
provided that the following conditions are met: 

\noindent
1). $\Delta \phi \sim 1$, i.e. the phase should be large enough.\\
2). $\phi (R = c t_u) < \Delta \phi$, where $t_u$ is the age of 
the Universe. This condition ensures the absence of the spatial 
averaging. \\ 
3). Events within  sufficiently small energy interval $\Delta E < E$ 
are selected to avoid the averaging over the energy. 

For $d = 50$ Mpc, $r_0 = 1$ Mpc we get $\Delta \phi/ \phi \simeq 2$. 
It means that for the supercluster the conditions 1). and 2). can 
be fulfilled for certain values of $E$ and $\Delta f$. 
However, observation of the asymmetry will require high statistics, 
since one should select the events within small solid angle,   
$\sim (r_0/d)^2$, and an energy-cut is required. 

The conditions 1). and 2). are not satisfied for our Galaxy 
as the source of the asymmetry. 

If an excess of $\nu_{\tau}$ events will not be found, then
this will allow one to exclude the region of parameters, 
$\sin^2 2\theta_g > 10^{-2}$ and $\Delta f > \Delta f_{AGN}$,  
where in the expression (\ref{fAGN}) for $\Delta f_{AGN}$ 
one should take as $\Delta m^2$ the upper experimental bound 
on neutrino mass difference. \\

As we have discussed in Sec. III most probably neutrinos are 
massive and mixed, and moreover, there is a good chance that 
forthcoming experiments will measure $\Delta m^2$ and 
$\sin^2 2\theta$. In the case of massive neutrinos a signature of 
the gravity effect is quite different from that of the massless case. 
One should look for the deviation of the observed ratio 
$\nu_{\tau} / \nu_{\mu}$'s (as well as $\nu_e/\nu_\mu$)   
from that stipulated by vacuum oscillations. 
The gravity induced mixing can both suppress and enhance 
$\nu_{\tau}$-signal.  

If $\Delta f$ is in Region II (\ref{fresonance}) 
the gravitational MSW effect occurs in the resonance channel. 
The resonance conversion of $\nu_\mu$ to $\nu_\tau$ gives rise 
to a large transition probability $P > 1/2$. Therefore, the 
observation of the ratio $\nu_{\tau} / \nu_{\mu}$'s larger than 
unity would provide an evidence for the gravitational MSW effect. 

In the nonresonant channel the gravity effect is described by 
the transition probability  (\ref{transition}) with minus sign. 
For $\theta_g > \theta$ ($\theta_g < \theta$), the gravity effect 
enhances (suppresses) the transition. The modification is maximal  
if the vacuum angle $\theta$ is small and the gravity angle 
is large, $\theta_g \sim \pi / 4$. 

If $\Delta f$ falls into Region III (\ref{gravdom}) the signal 
would mimic the one due to vacuum flavor oscillation. But, since 
we assume that the masses and vacuum mixing angles will be 
determined by future experiments, the difference between 
$\theta_g$ and $\theta$ should show up in the measured ratio 
of $\nu_\tau$ to $\nu_\mu$.\\

Let us discuss experimental signatures and estimate the sensitivity 
regions for the VEP parameters for two probable scenarios of neutrino 
masses and mixing. They are suggested by some experimental results at 
hand and will be checked by forthcoming experiments. 

\noindent
1. Suppose the heaviest neutrino (which practically coincides with
$\nu_{\tau}$) has the mass in the cosmologically interesting region:
$m_3 = (3 - 7)$ eV,  so that $\Delta m^2 = (10 - 50)$ eV$^2$
in the case  of mass hierarchy. For these values of mass the  
present experimental bound on mixing angle is already 
$\sin^2 2\theta < 5 \cdot 10^{-3}$,  i.e. below the sensitivity 
limit (\ref{sensitivity}). We will assume also that mixing with 
electron neutrinos is negligibly small.

In this case the effect of mass-induced vacuum oscillation can be 
neglected. The result for the ratio $\nu_{\tau}/\nu_{\mu} > 0.01$ 
in the underwater/ice detectors would then be an indicator of the 
VEP mechanism of flavor conversion. Observation of large value of 
the ratio:  $\nu_{\tau}/\nu_{\mu} \gsim 1$, would provide a clear
signature for the gravitational MSW effect. 
From the resonance condition we find, assuming the typical values 
for the parameters in (\ref{fAGN}) and (\ref{fIG}), the required 
region for $\Delta f$: 
\be
\label{delta}
\Delta f = (10^{-27} - 5\times 10^{-25})
\left(\frac{\Delta m^2}{10~ {\rm eV^2}}\right) \ . 
\ee
Then, for fixed $\Delta f$, the adiabaticity condition 
gives the bound on mixing angles  
\be
\label{admixing} 
(\tan2\theta - \tan2\theta_g)^2 \cos2\theta_g 
\gg 5 \times 10^{-9} 
\left(\frac{\Delta f}{5\times 10^{-25}} \right)^{-1}
\left(\frac{\Delta m^2}{10~ {\rm eV^2}}\right)^{-1}\ .
\ee
If lepton mixing is similar to the quark mixing, then one expects 
$\sin^2 2\theta > 10^{-3}$ for $\nu_{\tau} - \nu_{\mu}$. In this 
case the adiabaticity condition is satisfied by vacuum mixing alone 
and the gravitational angle can be arbitrarily small.   

Non-observation of such a signal will allow one to exclude the 
whole region of parameters in (\ref{delta}) and $\theta_g$ being 
not too close to $\theta$.   

Observation of a moderately-large ratio, 
$0.01 < \nu_{\tau}/\nu_{\mu} < 1$ would be the signal for 
one of the following three possibilities; 
the nonadiabatic gravitational MSW mechanism (in resonant channel), 
the nonresonant adiabatic conversion with transition probability 
$P_a \approx \sin^2 \theta_g$ (in nonresonant channel), 
or oscillations in the intergalactic gravitational field with 
$\sin^2 2\theta_g > 0.01$. In the first case, the parameter 
$\Delta f$ should take the value around (\ref{delta}) and 
$\theta_g \approx \theta$ to violate the adiabaticity which is 
well satisfied by vacuum mixing alone. In the other two cases, 
the region $\sin^2 2\theta_g \gsim 0.01$ will be probed.\\

\noindent
2. Suppose that the heaviest neutrino has a mass $m_3 \sim 0.1$ eV 
and $\nu_{\mu} - \nu_{\tau}$ mixing is large: 
$\sin^2 2\theta \sim 0.5 - 1$, so that $\nu_{\mu} - \nu_{\tau}$ 
oscillations solve the atmospheric neutrino problem. As the result 
of the vacuum oscillations one predicts the ratio 
$\nu_{\tau}/\nu_{\mu} \leq 1$ for neutrinos from AGN. 
The observation of larger ratio, $\nu_{\tau}/\nu_{\mu} \gsim 1$, 
will be a signature for the gravitational MSW effect. In this case 
the VEP parameter should be in the interval 
\be
\Delta f = (0.3 - 1) \times 10^{-30} - 5 \times 10^{-28} \ ,  
\ee 
assuming again the typical values of parameters in the parentheses in 
(\ref{fAGN}) and (\ref{fIG}). The adiabaticity condition is 
satisfied by large vacuum mixing angle, and the angle $\theta_g$ 
can be arbitrarily small. 

In the nonresonant channel the gravitational effect manifests 
in a different manner. If $\Delta f$ is in Region II 
(\ref{fresonance}), i.e., $\Delta f > 10^{-30}$, and the 
gravitational mixing angle $\theta_g$ is smaller than the vacuum 
angle $\theta$, the gravitational effect suppresses neutrino 
transition. Namely, the $\nu_\tau$ signal will be smaller than the 
one expected for mass-induced vacuum oscillations. 
Using  (\ref{transition}) with $\theta_g \ll 1$ we find 
transition probability $P = \sin^2 \theta$ instead of 
$P = \frac{1}{2} \sin^2 2\theta$ in  the absence of the VEP. 
In such a way the transition probability is reduced by 40\% and 
30\% for $\sin^2 2\theta = 0.6$ and 0.8, respectively.   

In Region III (\ref{gravdom})
the averaged transition probability equals  
$P = \frac {1}{2} \sin^2 2\theta_g$ 
and for small $\theta_g$ suppression can be much stronger, 
so that $\nu_{\tau}$ signal will not be observable at all.\\ 


\section{Conclusion} 

\noindent
1. Cosmological distances ( $\sim 100$ Mpc) and ultrahigh 
energies ($\sim 1$ PeV) of neutrinos from the AGN open 
unique possibility to improve an accuracy of testing the 
equivalence principle by 12 - 21 orders of magnitude. 

\noindent
2. For massless neutrinos VEP can induce the oscillations 
$\nu_{\mu} - \nu_{\tau}$ of cosmic neutrinos which may lead to 
observable $\nu_{\tau}$ signals in the large underwater/ice 
installations. The sensitivity to the parameters of the VEP can 
be estimated as $\Delta f \gsim 10^{-41}$ and 
$\sin^2 2\theta_g > 2 \cdot 10^{-2}$. 
In the case of the nonaveraged oscillations 
($\Delta f$ at the lower bound) one can expect  
an anisotropy of the $\nu_{\tau}$ signal correlated to 
the position of the Great Attractor. 

\noindent
3. In the case of massive neutrinos the gravitational effects due 
to VEP can modify the result of the vacuum oscillations. 
For certain values of the parameters neutrinos may undergo 
the resonance flavor conversion driven by the gravitational 
potential of AGN (or of our Galaxy, if the intergalactic potential 
is sufficiently weak). The gravitational effects become important if 
$\Delta f \gsim \Delta f_{AGN} \sim 10^{-28} (\Delta m^2 /1$ eV$^2$). 
For $\Delta f \sim  (1 - 10^3) \Delta f_{AGN}$ one may expect almost 
complete transition of $\nu_{\mu}$ to $\nu_{\tau}$ due to the 
resonance conversion. A strong observable effect may exist for 
arbitrarily small $\sin^2 2\theta_g$. 
For $\Delta f \gg  (1 - 10^3) \Delta f_{AGN}$ 
neutrinos undergo the gravity induced oscillations and vacuum 
mixing effect can be neglected. 

\noindent
4. The VEP effects can be identified if the gravitational mixing 
differs appreciably from vacuum mixing. In general, the VEP effect 
will manifest itself as deviation of the observed ratios    
$\nu_{\tau}/\nu_{\mu}$ and $\nu_{\mu}/\nu_{e}$ from those 
stipulated by vacuum oscillations. Of course, the latter can be 
predictable only if neutrino masses and mixing are determined by 
forthcoming neutrino experiments. For small vacuum mixing the VEP 
can enhance the $\nu_{\mu} - \nu_{\tau}$ transition and the 
$\nu_{\tau}$-signal. On the contrary, for large vacuum mixing 
the VEP can lead to suppression of the $\nu_{\tau}$-signal. 
The ratio $\nu_{\tau}/\nu_{\mu} > 1$ is the clear signature of 
the resonance flavor conversion. 

\noindent
5. If deviation from vacuum oscillation effects will not be found,  
then one will be able to exclude very large new region 
of the VEP - parameters.\\

\section{Acknowledgements}
The authors are grateful to Osamu Yasuda for useful discussions. 
One of us (A.S.) would like to thank Department of Physics, 
Tokyo Metropolitan University for hospitality. The other (H.M.) 
is supported in part by Grant-in-Aid for Scientific Research of 
the Ministry of Education, Science and Culture No. 0560355, 
and by Grant-in-Aid for Scientific Research under International
Scientific Research Program; Joint Research No. 07044092.


\begin{references}

\bibitem{AGN}
V. S. Berezinsky and V. L. Ginzburg, Mon. Not. Royal Ast. Soc. 
{\bf 194}, 3 (1981); 
F. W. Stecker, C. Done, M. H. Salamon, and P. Sommers,
Phys.\ Rev.\ Lett.\ {\bf 66}, 2697 (1991);
A. P. Szabo and R. J. Protheroe, Astroparticle Phys. {\bf 2}, 375 (1994).
See also T. Stanev, in {\it High Energy Neutrino Astrophysics}
(World Scientific Publishing Co., Singapore, 1992).

\bibitem{LP}
J. G. Learned and S. Pakvasa, Astroparticle Phys. {\bf 3}, 267 (1995).

\bibitem{ATM}
K. S. Hirata et al., Phys. Lett. {\bf B205}, 416 (1988); 
ibid {\bf B280}, 146 (1992);
R. Becker-Szendy et al., Phys. Rev. {\bf D46}, 3720 (1992);
Y. Fukuda et al., Phys. Lett. {\bf B335}, 237 (1994).

\bibitem{Gasp}
M. Gasperini, Phys.\ Rev.\ {\bf D38},  2635 (1988); {\bf D39},
3606 (1989).

\bibitem{HL}
A. Halprin and C. N. Leung, Phys.\ Rev.\ Lett.\ {\bf 47},
1833 (1991).

\bibitem{IMY}
K. Iida, H. Minakata and O. Yasuda, Mod.\ Phys.\ Lett.\ {\bf A8},
1037 (1993).

\bibitem{PHL}
J. Pantaleone, A. Halprin and C. N. Leung, Phys.\ Rev.\ {\bf D47},
R4199 (1993).

\bibitem{BNMB}
M. N. Butler, S. Nozawa, R. A. Malaney, and A. I. Boothroyed,
Phys.\ Rev.\ {\bf D47},   2615 (1993).

\bibitem{MN}
H. Minakata and H. Nunokawa, Phys.\ Rev.\ {\bf D51}, 6625 (1995).

\bibitem{BKL}
J. N. Bahcall, P. I. Krastev and C. N. Leung, Phys.\ Rev.\ {\bf D51},
1770 (1995).

\bibitem{mina}
H. Minakata, Nucl. Phys. B (Proc. Supple.) {\bf 38}, 303 (1995).

\bibitem{panta} A. Halprin, C. N. Leung and 
J. Pantaleone, UDHEP-11-95, UAAHEP-01-95. 

\bibitem{LKT}
M. J. Longo, Phys.\ Rev.\ Lett.\ {\bf 60}, 173 (1988);
L. M. Krauss and S. Tremaine, ibid.\ {\bf 60},  176 (1988).

\bibitem{LB}
D. Lynden-Bell et al., Astrophys. J. {\bf 326}, 19 (1988).

\bibitem{Kenyon}
The relevance of the supercluster to the gravitational effect in 
$K-\bar{K}$ oscillation has been pointed out by 
R. Kenyon, Phys. Lett. {\bf B237}, 274 (1990). See also, 
M. Good, Phys. Rev. {\bf 121}, 311 (1961).

\bibitem{Schmidt}
M. Schmidt, in {\it The Milky Way Galaxy}, IAU symposium No. 106 
(D. Reidel Publishing Co, Dordrecht, 1985). 

\bibitem{Xray}
R. F. Mushotzky, Astrophys. J. {\bf 256}, 92 (1982); 
T. J. Turner and K. A. Pounds, Mon. Not. Astr. Soc. 
{\bf 240}, 833 (1989). 

\bibitem{ABS}
R. Barbieri, J. Ellis and M. K. Gaillard, 
Phys. Lett., {\bf B90}, 249 (1980); 
E. Kh. Akhmedov, Z. G. Berezhiani and G. Senjanovi\'c, 
Phys. Rev. Lett., {\bf 69}, 3013 (1992). 

\bibitem{MSW}
 S. P. Mikheyev and A. Smirnov, Sov.\ J.\ Nucl.\ Phys.\ {\bf 42},
913 (1985);
L. Wolfenstein, Phys.\ Rev.\ {\bf D17}, 2369 (1978).

\end{references}
\end{document}